\documentstyle[12pt,epsfig]{article}
\begin{document}

\vskip 0.4cm
\vspace{1.2cm}
\begin{center}
\Large
\centerline{\bf Comment on the \lq\lq Coupling Constant and}
\vspace{3mm}
\centerline{\bf  Quark Loop Expansion for Corrections to}
\vspace{3mm}
\centerline{\bf the Valence Approximation'' by Lee and Weingarten}
\vspace{1.5cm}
\large

M. Boglione $^{1}$

\vspace{0.5cm}
and
\vspace{0.5cm}

\large

M.R. Pennington $^2$

\vspace{0.4cm}
\begin{em}
$^1$  Vrije Universiteit Amsterdam, De Boelelaan 1081,\\ 1081 HV Amsterdam, 
The Netherlands\\

\vspace{0.4cm}
$^2$ Centre for Particle Theory, University of Durham\\
     Durham DH1 3LE, U.K.\\

\end{em}

\end{center}
\normalsize
\baselineskip=7mm
\parskip=2mm
\vspace{5mm} 
\parbox{16 cm}{\leftskip=1cm\rightskip=1cm{
Lee and Weingarten have recently criticized our calculation of quarkonium 
and glueball scalars as being \lq\lq incomplete'' and \lq\lq incorrect''. 
Here we explain the relation of our calculations to full QCD.\\
}}

\centerline{PACS numbers: 14.40.Cs, 12.40.Yx, 12.39.Mk }

\centerline{Submitted to Physical Review 28th April 1999}
\vspace{1cm}

Lattice techniques provide an invaluable tool for calculating the
properties of hadrons~\cite{teper}.  As a matter of practical necessity, these
calculations involve approximations to full QCD.  While the
spectrum of glueballs has been computed with increasing precision~\cite{UKQCD,IBM1,MP}, this is
within quenched QCD. To make contact with experiment requires one to get
closer to the full theory by allowing for the creation of $q{\overline q}$
pairs. Different attempts to do this for light scalars, both quarkonium
and glueball, have been made by Boglione and Pennington (BP)~\cite{BP} and by
Lee and Weingarten (LW)~\cite{LW}. In a very recent paper, LW have criticized the
former attempt as being
{\it incomplete} and {\it incorrect}. We believe their extensive
discussion
is in error in claiming key aspects of  QCD have been omitted by BP. Let
us explain.

The BP treatment, like that of Tornqvist~\cite{nils} and others~\cite{vanb}, is based on a
specific approximation to QCD, in which only hadronic (color singlet)
bound states and their interactions occur.  One begins with the QCD
Lagrangian, for which the only parameters are quark masses and the
strength of the quark-gluon interaction. $\Lambda_{QCD}$, and other
scheme-dependent parameters, enter on renormalization. One then formally
integrates out the quark and gluon degrees of freedom and obtains a
Lagrangian involving only hadronic fields with their interactions, in an
infinite variety of ways, all of which are determined by the parameters of
the underlying theory. We then focus on the ten lightest scalar states.
The bare states are realized by switching off all their interactions. Consequently, their propagators are those of bare particles:
they are stable.  To take this limit, each coupling in the effective
Lagrangian of hadronic interactions is multiplied by a parameter
$\lambda_i$ and these $\lambda_i$ are taken to zero. This does not
necessarily correspond to a simple limit of QCD. Nevertheless, we plausibly
assume that the ten lightest non-interacting states, that result in this limit, are
the nine members of an ideally mixed quarkonium multiplet and an
(orthogonal) glueball. Notice that the names {\it quarkonium} and {\it
glueball} are just a convenient way of referring to the quantum numbers of
these states. Individual quark and gluon fields play no role. However,
they are, of course, implicit in the formation of hadronic bound states.

\begin{figure}[h]
\begin{center}
\vspace{0.5cm}
\mbox{~\epsfig{file=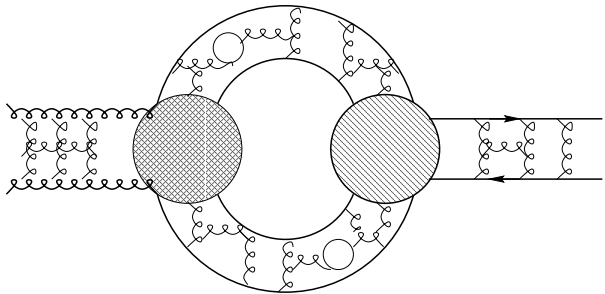,angle=0,width=8cm}}
\vspace{0.7cm}
\caption{ A pictorial representation of 
mixing between quarkonium and glueball physical states through common 
meson-meson channels, included in the Dyson summation of the scalar bound state propagator.}
\end{center}
\vspace{-3mm}
\end{figure}

The Tornqvist~\cite{nils} and BP treatment is then to switch on the \lq\lq dominant"
interactions of the light scalars by tuning the appropriate parameters
$\lambda_i$ from $0 \to 1$ for the couplings of the bound states
to two (or more) pseudoscalars \footnote{ for the glueball, the four pion
channel may be particularly important.}. It is by turning on the interactions 
that the bare states are ``dressed''. Fig.~1 represents the Dyson summation
of such contributions to the inverse propagator.
This dressing does not correspond
to the
creation of a single $q{\overline q}$ pair. Multiple pairs and all the
gluons (Fig.~1) needed to generate color singlets and respect the chiral limit 
are implicitly included.  
Indeed, it is well-known~\cite{sharpe}, that any picture of pions
as simple $q{\overline q}$ systems loses contact with the Goldstone nature
of the light pseudoscalars, so crucial for describing the world accessible
to experiment.  This important (chiral) limit is embodied in our
calculation. The resulting hadronic interactions have a dramatic effect on
the scalar sector. For instance, the $a_0$ and an $f_0$ emerge at 980 MeV
with large $K{\overline K}$ components~\cite{PDG}, even though their bare states are
members of an ideal multiplet 4-500 MeV/c$^2$ heavier.
LW criticize these results as not including the specific gluonic counterterm,
Fig.~2, and not explaining why. 

The explanation is clear~: our analysis
only includes color singlet states internally as
unitarity requires. Colored configurations of whatever kind are implicitly
included and not readily dissected. If such counterterms are relevant to
the dressing by pseudoscalar (Goldstone) pairs, they have been included.

%
\begin{figure}[h]
\begin{center}
\vspace{0.5cm}
\mbox{~\epsfig{file=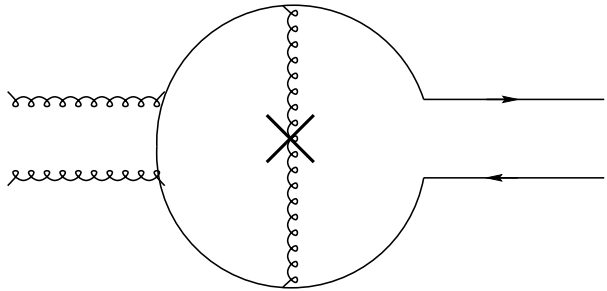,angle=0,width=8cm}}
\vspace{0.5cm}
\caption{Feynman diagram representing the counter term contribution to 
the glueball--quarkonium mixing amplitude as given by Lee and
 Weingarten \protect\cite{LW}.}
\end{center}
\vspace{-3mm}
\end{figure}
%

In spirit, our analysis~\cite{BP,BPnew} is close to that of Ref.~\cite{nils,vanb}. 
Propagators are dressed by hadron clouds, as in  Fig.~1. These determine the right hand cut structure of
meson-meson scattering amplitudes. However, in the work of Refs.~\cite{nils},
 this $s$--channel dynamics is assumed to control
the whole scattering amplitude, with left hand cut effects 
(and crossed-channel exchanges) neglected, even though this violates crossing
symmetry~\cite{Speth,BPnew}.
In our treatment~\cite{BP,BPnew}, particularly here where we consider mixing, 
only propagators are computed and no further assumptions are needed.

 Of course, our analysis does have approximations. For instance, the scale
of hadronic form-factors for a gluish state is assumed to be similar to
that of well-established $q{\overline q}$ hadrons. This may not be the
case.  Moreover, our treatment only incorporates  interactions with two
pseudoscalars, and to a lesser extent with multipion channels. It is these
that determine both the sign and magnitude of the mass-shifts generated.
For the quarkonium states, the dressing by the light two pseudoscalar
channels always produces a downward shift in mass. The size of these
shifts of between one and five hundred MeV (depending on flavor) is set
phenomenologically~\cite{PDG} by the $K^*_0(1430)$. A much smaller shift of 10--25
MeV
for the precursor glueball is set by the strength of the glueball to two
pseudoscalar coupling calculated on the lattice by Sexton {\it et al.}~\cite{gPP}. The suppression
of the couplings of the resulting \lq\lq dressed" hadron 
to two pseudoscalars happens~\cite{BPnew} irrespective
of the exact mass of the bare glueball~\cite{UKQCD,IBM1,MP}. The inclusion of
more channels, like $\rho\rho$ and $K^* {\overline {K^*}}$, may well be
important in dressing this state and alter the rather small mass-shifts
we found for that sector both in magnitude and sign. Of course, only
physically accessible hadronic intermediate states contribute to the
imaginary part of the propagator, Fig.~1. Unopen channels
contribute only to the real (or dispersive) part and result in
renormalizations of the undressed parameters. 

By including, in our calculation key aspects of the
hadron world, in the way described here and in ~\cite{BP}, we believe we must
have approached closer to full QCD --- despite the criticism of Lee and
Weingarten.

\vspace{5mm}

\noindent{\bf Note added}: Long after the submission of this paper to Physical Review,
Lee, Vaccarino and Weingarten have repeated their comments identically in Refs.~\cite{newwein1,newwein2}.
\vspace{5mm}

\noindent The authors acknowledge partial support 
from the EU-TMR Program {\it EuroDA$\Phi$NE}, Contract No.
CT98-0169.

\end{document}